\documentclass[unsortedaddress,preprint,aps,showpacs]{revtex4}
\usepackage{graphicx}
\usepackage{dcolumn}
\usepackage{amsmath}
\usepackage{amssymb}
\usepackage{amsfonts}
\usepackage{bm}
\usepackage{color}
\newcommand{\be}{\begin{equation}}
\newcommand{\ee}{\end{equation}}
\newcommand{\ba}{\begin{eqnarray}}
\newcommand{\ea}{\end{eqnarray}}
\begin{document}
\title{Shape and area fluctuation effects on nucleation theory}
\author{Santi Prestipino$^1$\footnote{Corresponding author. E-mail:
{\tt sprestipino@unime.it}}, Alessandro Laio$^2$\footnote{E-mail:
{\tt laio@sissa.it}}, and Erio Tosatti$^{2,3}$\footnote{E-mail:
{\tt tosatti@sissa.it}}}
\affiliation{
$^1$Universit\`a degli Studi di Messina, Dipartimento di Fisica e di Scienze 
della Terra, Contrada Papardo, I-98166 Messina, Italy \\
$^2$International School for Advanced Studies (SISSA) and UOS 
Democritos, CNR-IOM, Via Bonomea 265, I-34136 Trieste, Italy \\
$^3$The Abdus Salam International Centre for Theoretical Physics (ICTP),
P.O. Box 586, I-34151 Trieste, Italy}
\date{\today}
\begin{abstract}
In standard nucleation theory, the nucleation process is characterized by
computing $\Delta\Omega(V)$, the reversible work required to form a cluster
of volume $V$ of the stable phase inside the metastable mother phase. 
However, other quantities besides the volume could play a role in the free 
energy of cluster formation, and this will in turn affect the nucleation 
barrier and the shape of the nucleus. Here we exploit our recently introduced 
mesoscopic theory of nucleation to compute the free energy cost of a
nearly-spherical cluster of volume $V$ and a fluctuating 
surface area $A$, whereby
the maximum of $\Delta\Omega(V)$ is replaced by a saddle point in
$\Delta\Omega(V,A)$. Compared to the simpler theory based on volume only,
the barrier height of $\Delta\Omega(V,A)$ at the transition state is
systematically larger by a few $k_BT$. More importantly, we show that, 
depending on the physical situation, the most probable shape of the
nucleus may be highly non spherical, even when the surface tension and
stiffness of the model are isotropic.
Interestingly, these shape fluctuations do not influence or modify the
standard Classical Nucleation Theory manner of extracting the interface 
tension from the logarithm of the nucleation rate near coexistence. 
\end{abstract}
\maketitle

\section{Introduction}
\setcounter{equation}{0}
\renewcommand{\theequation}{1.\arabic{equation}}

When in a first-order phase transition a thermodynamic phase turns
metastable, it may remain stuck
for long in a state of apparent equilibrium until a favorable fluctuation
triggers the formation of the truly stable phase. Nucleation concerns the 
early stages of the phase transformation, which initially occurs as
an activated process~\cite{Kashchiev}. Despite many attempts to 
formulate a quantitatively accurate theory of homogeneous nucleation, 
the important problem of relating the nucleation rate (the main 
experimentally accessible quantity) to the microscopic features of the 
system still remains open. A less ambitious program is
to find a simple statistical model where a number of nucleation-related 
issues can find at least a partial answer. In a pair of recent 
papers~\cite{Prestipino1,Prestipino2}, we focused on a mesoscopic scale
model of this sort, in the form of a field theory in the surface of the 
nucleating cluster. While the classical nucleation theory (CNT) envisages 
the cluster surface as sharp and spherical with the same interface
free energy as the bulk-coexistence interface, the cluster of our
theory can make excursions around a reference shape, with a cost expressed
in terms of the parameters of a Landau free energy. Within this theory, two 
main results were obtained: i) The cluster formation energy shows, in 
addition to Landau-type corrections reflecting the finite width of the cluster 
interface~\cite{Fisher}, a term logarithmic in the cluster volume $V$,
with a numerical prefactor whose magnitude {\it and sign} are
only sensitive to the extent of interface anisotropy;
ii) The subleading corrections to the CNT free energy can so 
much affect the steady-state nucleation rate that the customary way of 
extracting the interface tension from it, based on the standard CNT 
recipe may easily lead to wrong results.

Here we pose another question, and give a detailed answer still in terms 
of our theory, 
concerning the role of the {\it area} $A$ of the nucleation cluster.  
A central assumption in CNT is that a single reaction coordinate
(the cluster size or volume) is sufficient to describe the nucleation cluster. 
This is so frequently and commonly adopted
that it is not always appreciated 
that such a hypothesis is actually only a convenient 
approximation. To be sure, there exist many notable exceptions.
In Refs.\,\cite{Pan,Moroni,Trudu,Peters,Zykova-Timan,Lechner,Russo}
microscopic attempts were
described that go beyond a single reaction coordinate, with important 
additional insights into the actual mechanism of nucleation.
In atomistic simulations in particular,
nucleation can be monitored by means of convenient order parameters
and the nucleation landscape can be
mapped out in terms of these variables. A general finding 
is the extreme irregularity of cluster shapes which, generally far from
spherical, are neither compact nor necessarily one-phase 
objects. However, atomistic studies are numerical in nature, 
and therefore intrinsically system-specific.

In this paper, we base on a generic field theory description
a study of the modifications in the energetics of nucleation
when, besides the cluster volume, the surface area is introduced
as a reaction variable.
Notwithstanding the simpler and necessarily more abstract nature
of our approach compared with atomistic ones, we show that this
additional variable, the area, is in many cases irrelevant
for the nucleation process, but becomes important when
the activation barrier to nucleation is small.
The instantaneous and average surface area of the nucleus are
significantly larger than that of the sphere of same volume. Moreover,
the free-energy barrier corresponding to the nucleation process
is systematically underestimated if one considers only
the volume as 
a reaction variable. We also provide a quantitative estimate of these effects
as a function of the model parameters, and inquire whether the standard
CNT procedure of extracting the interface free energy
from the logarithm of the nucleation rate is going to be affected by 
an average cluster area larger than spherical.

The paper is organized as follows. In Sections II and III we briefly 
recollect the features and main results of the field theory at the basis of 
our calculations. Next, in Section IV we present data for the nucleation
landscape as a function of volume and area of the cluster. The dependence
of the critical size and the barrier height on the model parameters are
investigated in detail. In Section V, we address
the issue how to extract the
interface tension from the measured nucleation rate in the light of
our new results. Final remarks and conclusions are given in Section VI.

\section{Review of the model}
\setcounter{equation}{0}
\renewcommand{\theequation}{2.\arabic{equation}}

In Refs.\,\cite{Prestipino1,Prestipino2} we introduced a
model description of the free energy of a homogeneous nucleation
cluster as a function of the cluster volume $V$.
The theory goes beyond CNT, in that it allows for 
fluctuations of the cluster surface $\Sigma$ around its mean shape. 
Two cases were considered, both amenable to
analytic treatment.
A quasispherical cluster, corresponding to an isotropic interface, and a
cuboidal cluster, addressing the opposite limit 
of strongly anisotropic interface tension.
We make use of the same theory here, to address the area dependence
of the cluster-free energy cost of a nearly-spherical cluster.

We first introduce the relevant thermodynamic framework,
slightly deviating from the notation used in \cite{Prestipino1,Prestipino2}.
Let the metastable and stable phases be called, respectively, 1 and
2 (for instance, supercooled liquid and solid close to melting).
If the basic variable, or reaction coordinate, is chosen to be the
volume $\cal V$ of the phase 2 cluster, then the
external control parameters are the temperature $T$, 
the volume $V_{\rm tot}$ of the vessel, and the chemical potential $\mu$. 
Let further $P_1$ and $P_2$ be the equilibrium 
pressure values in the two infinite 
phases for the given $T$ and $\mu$ values. For example, slightly below
the coexistence temperature $T_m$ and for $\mu=\mu_m$,
the chemical potential value at coexistence,
$\Delta P\equiv P_2-P_1$ is roughly equal to
$-L_m/(V_{\rm tot}T_m)\Delta T$, where $\Delta T=T-T_m$ and $L_m$ is 
the heat of fusion. In a long-lived metastable 1 state not far from 
coexistence, shape fluctuations of the 1-2 interface in a cluster of phase 2
occur with a weight proportional to the Boltzmann factor relative to a 
coarse-grained Hamiltonian ${\cal H}[\Sigma]$ (here, a Landau grand 
potential), given by
\be
{\cal H}[\Sigma]=-P_1(V_{\rm tot}-{\cal V}[\Sigma])-P_2{\cal V}[\Sigma]+{\cal H}_s[\Sigma]\,,
\label{2-1}
\ee
where ${\cal V}[\Sigma]$ is the cluster volume enclosed by $\Sigma$ and
${\cal H}_s[\Sigma]$ is the free-energy functional accounting for the cost 
of the interface (note that, at this level of generality, it is not even 
necessary that $\Sigma$ be a connected surface). 

For the ${\cal H}_s[\Sigma]$ in Eq.\,(\ref{2-1}) we assume a Canham-Helfrich 
form, containing spontaneous-curvature and bending-energy terms in 
addition to interface tension, with parameters derived from a more 
fundamental Landau free energy. In detail, denoting by $H$ the 
mean curvature of the surface $\Sigma$ of the cluster, the interface
free-energy functional reads
\be
{\cal H}_s[\Sigma]=\int_\Sigma{\rm d}S\,\left(\sigma_m-2\sigma_m\delta_m H+2\lambda H^2\right)\,,
\label{2-2}
\ee
where the system-specific quantities $\sigma_m,\delta_m$, and
$\lambda$ would generally depend on the local surface orientation (see the 
form of these coefficients in \cite{Prestipino2}).

As mentioned above, two limiting cases of Eq.\,(\ref{2-2}) can be studied 
analytically, those of isotropic and of extremely anisotropic interfaces.
In the isotropic case, $\sigma_m,\delta_m$, and $\lambda$ are constant 
parameters and the shape of the cluster is on average spherical. Although 
the solid-liquid interface is notoriously anisotropic, in many cases (hard 
spheres, Lennard-Jones fluid, etc.) the anisotropy is small enough to be 
neglected as a first step. When deviations from sphericity
are small, the equation for $\Sigma$ can be expressed in spherical 
coordinates as $R(\theta,\phi)=R_0[1+\epsilon(\theta,\phi)]$ with
$\epsilon(\theta,\phi)\ll 1$. Denoting by $x_{l,m}$ the Fourier coefficients
of $\epsilon(\theta,\phi)$ on the basis of real spherical harmonics, and 
discarding terms of order higher than the second in these coefficients, the 
functional ${\cal H}_s$ takes the explicit form~\cite{Prestipino2}:
\ba
&& {\cal H}_s=4\pi\sigma_m R_0^2+\frac{\sigma_m R_0^2}{2}\sum_{l>0,m}(l^2+l+2)x_{l,m}^2-8\pi\sigma_m\delta_m R_0
\nonumber \\
&& -\sigma_m\delta_m R_0\sum_{l>0,m}l(l+1)x_{l,m}^2+8\pi\lambda+\frac{\lambda}{2}\sum_{l>1,m}l(l+1)(l-1)(l+2)x_{l,m}^2\,.
\label{2-3}
\ea

\section{Nucleation cluster of volume V and area A: the restricted grand potential}
\setcounter{equation}{0}
\renewcommand{\theequation}{3.\arabic{equation}}

The model described by Eqs.\,(\ref{2-1}) and (\ref{2-2}) assumes
that the relevant collective variable (CV) for describing the process is
the volume $V$ of the nascent cluster. 
Under this assumption, the relevant thermodynamic potential is the restricted
grand potential for a predominantly $1$ system with an inclusion of phase 2 of 
arbitrary shape but fixed volume $V$:
\be
\Omega_{1+2}(V)=-\frac{1}{\beta}\ln\left\{a^3\int{\cal D}\Sigma\,\delta({\cal V}[\Sigma]-V)e^{-\beta{\cal H}[\Sigma]}\right\}\,,
\label{3-1}
\ee
where, on the right-hand side, $\beta=(k_BT)^{-1}$. In Eq.\,(\ref{3-1}), $a$ is 
a microscopic length and ${\cal D}\Sigma$ is a dimensionless integral
measure. For the same choice of eigenfunctions as in Eq.\,(\ref{2-3})
the integral measure reads~\cite{Prestipino2}
\be
\int{\cal D}\Sigma=\int_{-\infty}^{+\infty}\prod_{l>0,m}\left(\frac{S}{s}\,{\rm d}x_{l,m}\right)\int_0^{+\infty}\frac{{\rm d}R_0}{a}\,,
\label{3-2}
\ee
where $S=(36\pi)^{1/3}V^{2/3}$ is the area of the spherical surface of 
volume $V$ and $s=4\pi a^2$. We emphasize that, due to the existence 
of a lower cutoff of $a$ on interparticle distances, a $l$ upper cutoff of 
$l_{\rm max}=\sqrt{S}/a-1$ is implicit in Eq.\,(\ref{3-2}). 
Hence, $V$ cannot take any values but only those related to
$l_{\rm max}$ via
\be
(36\pi)^{1/3}V^{2/3}=S=(l_{\rm max}+1)^2a^2\,,\,\,\,\,\,\,l_{\rm max}=2,3,4,\ldots
\label{3-3}
\ee

The grand potential of $1$ is 
simply $\Omega_1=-P_1 V_{\rm tot}$, although a different but 
equivalent expression is also possible, considering that, by its very 
nature, phase 1 contains small clusters of phase 2 in its interior.
Denoting $V_{\rm max}$ the maximum volume an inclusion of 2
can have without altering the nature of 1, we can also write
\be
\Omega_1=-\frac{1}{\beta}\ln\int_{{\cal V}[\Sigma]<V_{\rm max}}{\cal D}\Sigma\,e^{-\beta{\cal H}[\Sigma]}
\label{3-4}
\ee
(the value of $V_{\rm max}$ is close above the critical volume $V^*$, 
i.e., the volume in the transition state).

The grand-potential excess $\Delta\Omega(V)$, providing the
reversible/minimum work needed to form
a 2-phase inclusion of volume $V$ within 1, is evaluated as
\ba
\Delta\Omega(V)\equiv\Omega_{1+2}(V)-\Omega_1 &=& (P_1-P_2)V-\frac{1}{\beta}\ln\left\{a^3\int{\cal D}\Sigma\,\delta({\cal V}[\Sigma]-V)e^{-\beta{\cal H}_s[\Sigma]}\right\}
\nonumber \\
&\equiv& -V\Delta P+F_s(V)\,,
\label{3-5}
\ea
$F_s(V)$ being the surface free energy. Equation (\ref{3-5}) resembles 
the free-energy barrier of CNT, with the key difference that the CNT cost 
for the surface is only the leading term in $F_s(V)$. Finally, there is a 
simple relation between $\Delta\Omega(V)$ and the probability density of 
volume, defined as
\be
\rho(V)\equiv\frac{\int{\cal D}\Sigma\,\delta({\cal V}[\Sigma]-V)\exp\{-\beta{\cal H}[\Sigma]\}}{\int_{{\cal V}[\Sigma]<V_{\rm max}}{\cal D}\Sigma\,\exp\{-\beta{\cal H}[\Sigma]\}}\,.
\label{3-6}
\ee
Using Eqs.\,(\ref{3-1}) and (\ref{3-4}), it promptly follows that
\be
-\frac{1}{\beta}\ln\left\{\rho(V)a^3\right\}=\Delta\Omega(V)\,,
\label{3-7}
\ee
which provides a way to calculate $\Delta\Omega$
numerically~\cite{tenWolde,Bowles,Maibaum}. Simulations show that, 
unless $V$ is very small, an overwhelming fraction of 2 particles
is gathered in a single cluster, as indeed expected from the arguments in 
\cite{tenWolde2}. A connected 2-phase inclusion within 1 is also a 
leading assumption of the theory of Refs.\,\cite{Prestipino1,Prestipino2}.

With these stipulations,
the free-energy cost of cluster formation for large $V$ turns out to be
\be
\Delta\Omega(V)=-V\Delta P+\widetilde{A}\,V^{2/3}+\widetilde{B}\,V^{1/3}+\widetilde{C}-\frac{7}{9}k_BT\ln\frac{V}{a^3}\,,
\label{3-8}
\ee
with $\widetilde{A},\widetilde{B},\widetilde{C}$ explicit functions of
$\sigma_m,\delta_m$, and $\lambda$ given in Ref~\cite{Prestipino2}.
Equation (\ref{3-8}) represents a step forward from CNT, as confirmed by 
explicit simulations in the Ising model~\cite{Prestipino1,Prestipino2}. 

Here we proceed to characterize the quasispherical cluster
by means of a coarse-grained
free energy function where, besides the volume,
we use the area $A$ of the cluster surface as a second CV:
\ba
\Delta\Omega(V,A) &=& -k_BT\ln\left\{a^5\int{\cal D}\Sigma\,\delta({\cal V}[\Sigma]-V)\delta({\cal A}[\Sigma]-A)e^{-\beta{\cal H}[\Sigma]}\right\}
\nonumber \\
&\equiv& -V\Delta P+F_s(V,A)\,.
\label{3-9}
\ea
The meaning of $\Delta\Omega(V,A)$ is the cost of forming a solid cluster 
of area $A$ and volume $V$ out of the liquid. The last term in (\ref{3-9})
(i.e., the surface free energy) is given by:
\be
e^{-\beta F_s(V,A)}=a^5\int{\cal D}\Sigma\,\delta({\cal V}[\Sigma]-V)\delta({\cal A}[\Sigma]-A)e^{-\beta{\cal H}_s}\,,
\label{3-10}
\ee
and the following sum rule holds:
\be
\int_0^{+\infty}\frac{{\rm d}A}{a^2}\,e^{-\beta\Delta\Omega(V,A)}=e^{-\beta\Delta\Omega(V)}\,,
\label{3-11}
\ee
which provides a useful consistency check of the calculation.

We proceed as for the earlier computation of $\Delta\Omega(V)$ in 
\cite{Prestipino2}, by first carrying out the trivial integral over $R_0$.
The result is:
\ba
&& e^{-\beta F_s(V,A)}=(36\pi)^{-2/3}\left(\frac{V}{a^3}\right)^{-4/3}e^{-8\pi\beta\lambda}e^{8\pi\beta\sigma_m\delta_m\left(3V/(4\pi)\right)^{1/3}}e^{-\beta\sigma_m A}
\nonumber \\
&\times& \int_{-\infty}^{+\infty}\prod_{l>0,m}\left(\frac{S}{s}{\rm d}x_{l,m}\right)
\exp\left(-\frac{1}{4\pi}\sum_{l>0,m}x_{l,m}^2\right)
\nonumber \\
&& \times\exp\left(-\frac{\beta\lambda}{2}\sum_{l>1,m}l(l+1)(l-1)(l+2)x_{l,m}^2\right)
\nonumber \\
&& \times\exp\left(\beta\sigma_m\delta_m\left(\frac{S}{4\pi}\right)^{1/2}\sum_{l>1,m}(l^2+l-2)x_{l,m}^2\right)
\nonumber \\
&& \times\delta\left(1+\frac{1}{8\pi}\sum_{l>1,m}(l^2+l-2)x_{l,m}^2-(36\pi)^{-1/3}V^{-2/3}A\right)\,.
\label{3-12}
\ea
Note that the delta-function argument is strictly positive for
$A<(36\pi)^{1/3}V^{2/3}$, yielding in this case $F_s(V,A)=+\infty$.
This just expresses the well-known fact that the sphere has the smallest 
surface area among all surfaces enclosing a given volume. Hence, we 
take $A>(36\pi)^{1/3}V^{2/3}$ in the following and define the
deviation from sphericity as
\be
\alpha\equiv(36\pi)^{-1/3}V^{-2/3}A-1>0\,.
\label{3-13}
\ee
Using the integral representation of the delta function, we obtain:
\ba
&& e^{-\beta F_s(V,A)}=(36\pi)^{-2/3}\left(\frac{V}{a^3}\right)^{-4/3}e^{-8\pi\beta\lambda}e^{8\pi\beta\sigma_m\delta_m\left(3V/(4\pi)\right)^{1/3}}e^{-\beta\sigma_m A}\frac{1}{2\pi}\int_{-\infty}^{+\infty}{\rm d}k\,e^{-i\alpha k}
\nonumber \\
&\times& \int_{-\infty}^{+\infty}\prod_{l>0,m}\left(\frac{S}{s}{\rm d}x_{l,m}\right)\exp\left\{-\frac{1}{4\pi}\sum_{l>0,m}\left[1+2\pi\beta\lambda\,l(l+1)(l-1)(l+2)\right.\right.
\nonumber \\
&& \left.\left.-4\pi\beta\sigma_m\delta_m\left(\frac{S}{4\pi}\right)^{1/2}(l^2+l-2)
-\frac{i}{2}(l^2+l-2)k\right]x_{l,m}^2\right\}
\nonumber \\
&=& (36\pi)^{-2/3}\left(\frac{V}{a^3}\right)^{-4/3}e^{-8\pi\beta\lambda}e^{8\pi\beta\sigma_m\delta_m\left(3V/(4\pi)\right)^{1/3}}e^{-\beta\sigma_m A}\left(\frac{2\pi S}{s}\right)^{\sum_{l=1}^{l_{\rm max}}(2l+1)}
\nonumber \\
&\times& \frac{1}{2\pi}\int_{-\infty}^{+\infty}{\rm d}x\,\frac{e^{-i\alpha x}}{\prod_{l=2}^{l_{\rm max}}\left\{c_l(S)-\frac{i}{2}(l^2+l-2)x\right\}^{(2l+1)/2}}\,,
\label{3-14}
\ea
where
\be
c_l(S)=1+2\pi\beta\lambda\,l(l+1)(l-1)(l+2)-4\pi\beta\sigma_m\delta_m(l^2+l-2)\left(\frac{S}{4\pi}\right)^{1/2}\,.
\label{3-15}
\ee
The last step in (3.14) is only justified when all $c_l(S)>0$.
A problem then occurs for $\delta_m>0$ since, above a certain value
of the volume, $c_2(S)$ becomes negative and $F_s(V,A)$
ceases to be defined.
For $l_{\rm max}=2$, the integral in (\ref{3-14}) can be evaluated 
analytically (see appendix A). In the other cases, this integral is best 
converted into a real integral,
\ba
&& \int_{-\infty}^{+\infty}{\rm d}x\,\frac{e^{-i\alpha x}}{\prod_{l=2}^{l_{\rm max}}\left\{
c_l(S)-\frac{i}{2}(l^2+l-2)x\right\}^{(2l+1)/2}}
\nonumber \\
&=& 2\prod_{l=2}^{l_{\rm max}}c_l(S)^{-(2l+1)/2}\int_0^{+\infty}{\rm d}x\,\frac{\cos\left[\alpha x-\sum_{l=2}^{l_{\rm max}}\frac{2l+1}{2}\arctan\left(\frac{l^2+l-2}{2}\frac{x}{c_l(S)}\right)\right]}{\prod_{l=2}^{l_{\rm max}}\left[1+\left(\frac{l^2+l-2}{2}\frac{x}{c_l(S)}\right)^2\right]^{(2l+1)/4}}\,,
\label{3-16}
\ea
which is easier to compute numerically. We used Eq.\,(\ref{3-16})
to evaluate $\Delta\Omega(V,A)$ up to $l_{\rm max}=14$ for a number
of combinations of the model parameters.

\section{Results}
\setcounter{equation}{0}
\renewcommand{\theequation}{4.\arabic{equation}}

In Fig.\,1 we plot the contour lines of $\Delta\Omega(V,A)$ in the
$(V,\alpha)$ plane for a specific yet arbitrary choice of model parameters.
A clear saddle point is seen on the free-energy surface, marked by an
asterisk in the lower panel of Fig.\,1. The transition state for nucleation
is nothing but this free-energy saddle, which is the
``mountain pass'' separating the basin of attraction of the liquid ($V=0$)
from the region of $(V,A)$ points which, under the system
dynamics, would flow downhill to the ``solid'' sink at $V=+\infty$.

%
%
\begin{figure}
\centering
\includegraphics[width=11cm]{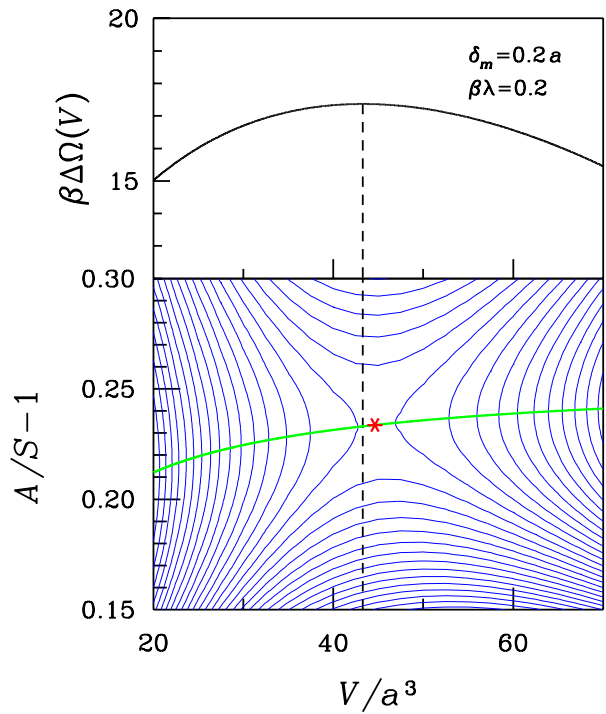}
\caption{(Color online). Quasispherical cluster: $\Delta\Omega(V)$ (top)
and $\Delta\Omega(V,A)$ (bottom) in units of $k_BT$, for a specific set
of model parameters
($\beta\Delta P\,a^3=1,\beta\sigma_ma^2=1,\delta_m=0.2\,a$, and
$\beta\lambda=0.2$). For these as well as other values of the parameters
we have checked by visual inspection that $\Delta\Omega(V,A)$ is indeed a
concave function of $V$ and a convex function of $A$ (and of $\alpha$ as well).
To have a better view of the figure, the two-dimensional nucleation landscape
has been represented through the contour lines of $\Delta\Omega(V,A)$ in the
$(V,\alpha)$ plane.
The green solid line in the bottom panel marks the minimum-free-energy
path $\alpha_{\rm min}(V)$.
The red asterisk marks the position of the saddle point of $\Delta\Omega(V,A)$
as computed through the interpolation procedure outlined in the text.
In the case considered, the critical volume increases by roughly
$3\%$ when the second collective variable $A$ 
is introduced, whereas the barrier height changes
from 17.368 to 19.343 ($+11\%$).
We checked numerically for $l_{\rm max}=3,4,5$ that
Eq.\,(\ref{3-11}) is exactly fulfilled (for $l_{\rm max}=2$ this
is done analytically in appendix A).}
\label{fig1}
\end{figure}

When averaged over many different dynamical trajectories, the nucleation
process can be described as following the lowest-free-energy route since
the Boltzmann weight is highest at the bottom of the free-energy valley.
However, due to the statistical nature of nucleation, individual
nucleation events also involve
some excursions up the walls of the valley, which are more frequent on
the high-$\alpha$ side because of the far more numerous shapes
available there for the cluster. In particular, uphill excursions on the
free-energy surface away from the saddle point along the $A$ direction
provide the cost of fluctuations of the nucleus about its mean shape.
Clearly, both the most favorable
nucleation pathway as well as the extent of corrugations of the nucleus
surface above its mean shape vary with the theory parameters. For
the case reported in Fig.\,1 (and in many other cases as well) the value
of $\alpha$ along the minimum-free-energy path increases
very slowly with $V$, apparently approaching a finite value at infinity.

A non-zero saddle-point value of $\alpha$ implies that the nucleus
-- which is spherical only on average -- has ripples in its surface.
This is not particularly surprising, considering that it is convenient
for the cluster to deviate from perfect sphericity in order to gain entropy
from shape fluctuations -- a finite-size roughening.  
The perfect sphere exerts an entropic
repulsion on the cluster shape, which is similar to the mechanism at the
origin of the free wandering of an interface away from an attractive hard
wall above the depinning temperature~\cite{Burkhardt}.

In order to calculate the saddle-point coordinates $(V^*,\alpha^*)$
for given values of the
parameters, we first computed the minimum of $\Delta\Omega$ as a function
of $\alpha$ for each $l_{\rm max}$; then, after extending $l_{\rm max}$
to a continuous variable, we maximized $\Delta\Omega$ along
the lowest-free-energy route just determined 
and eventually converted the
result in $V$ units. In a few cases, including the example in Fig.\,1,
we checked that this procedure gives exactly the same saddle
point as revealed by the contour plot.

%
%
\begin{figure}
\centering
\includegraphics[width=11cm]{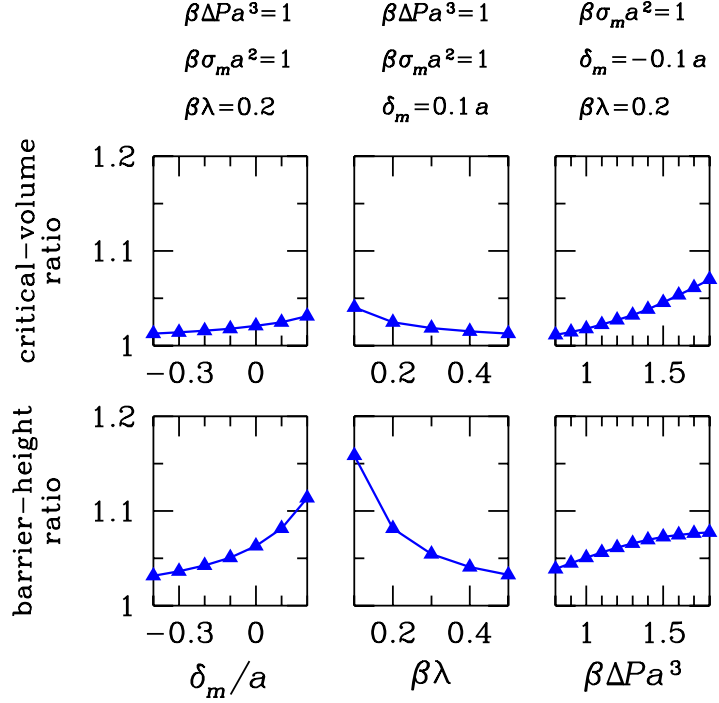}
\caption{(Color online).
Top: ratio between values of the critical volume $V^*$ obtained
with $(V,A)$ as collective variables and those obtained with $V$ only,
plotted as a function of the three model parameters $\delta_m,\lambda$,
and $\Delta P$ -- one at a time. Bottom: same ratio, now between
values of the barrier height $\Delta\Omega^*$. The critical volume and
the barrier height are both systematically larger in the $(V,A)$ case.
Observe that, for $\beta\lambda=0.2$ and $\delta_m=0.3\,a$ ($0.4\,a$),
the maximum value of $l_{\rm max}$ for which the integral in (3.14) still
converges is 6 (respectively, 4), i.e., too low to identify a saddle point on
$\Delta\Omega(V,A)$.
}
\label{fig2}
\end{figure}

The main message from Fig.\,1 is that the critical volume $V^*$ is larger
when allowing for two CVs, $(V,A)$, than for $V$ only.
The same holds for $\Delta\Omega^*$. The latter result 
is true in general as is seen in Fig.\,2,
which reports one- and two-CV values of $V^*$ and $\Delta\Omega^*$
in a wide range of $\delta_m,\lambda$, and $\Delta P$.
The underlying reason is that the non-linear procedure of obtaining
$\Delta\Omega(V)$ from
$\Delta\Omega(V,A)$ by integrating out the $A$ variable (Eq.\,(\ref{3-11}))
unavoidably corrupts the critical volume and the barrier height
causing both to appear artificially smaller than their true value, 
unless the minimum free-energy path were exceptionally
parallel to the $V$ axis.
The impact on $V^*$ and $\Delta\Omega^*$ of 
treating area as a collective variable 
besides volume is stronger when the barrier is low,
leading to barrier-height increases as large as $15\%$
in the cases plotted (but twice as that for e.g.
$\delta_m=0.1\,a,\beta\lambda=0.1$, and $\beta\Delta Pa^ 3=1.5$).
On the other hand, in most cases
the relative changes of $V^*$ and $\Delta\Omega^*$ are only a few
percent. This could explain why, in simulations of the
Ising model~\cite{Pan}, cluster area was found to play only a minor role
in the dynamics of nucleation.
As a side note, we observe that the $\Delta P$ value at which
$\Delta\Omega^*$ would extrapolate to zero is larger in the
two-CV case. This suggests that the spinodal threshold is
always underestimated in a treatment where only one reaction
variable ($V$) is considered.

Looking at Fig.\,2, we see that the behavior of $V^*$ and
$\Delta\Omega^*$ is similar. They both increase with reducing $\delta_m$
and with increasing $\lambda$, as may be expected from the form
(\ref{2-2}) of the interface free-energy functional, which shows that
in general a larger cost should be paid for the interface when
$-\delta_m$ and $\lambda$ are larger.

As coexistence is approached, the nucleus becomes effectively
flatter, since the mean radial amplitude of the surface ripples,
growing as $\sqrt{\ln(V^*/a^3)}$ as expected for a thermodynamically
rough interface, becomes negligible in comparison with the
nucleus radius. Despite that, it is not a priori clear what the
critical area ratio $\alpha^*$ should do in the coexistence limit,
where the critical nucleus volume diverges. Upon plotting
$\alpha^*$ as a function of supersaturation for fixed values of the
other parameters, we see that $\alpha^*$ increases slowly as $\Delta P$
is reduced (see Fig.\,3), apparently saturating to approach 
a finite value at
coexistence. Hence, we conclude that the weak -- even if unlimited --
growth of the interface width with volume yields a quantitatively
modest residual corrugation of the nucleation cluster which is unable
to change the scaling of cluster area from $V^{2/3}$ to a higher power,
and apparently 
even to a marginally faster increase such as $V^{2/3}\ln(V/a^3)$.
This expectation finds a confirmation in appendix B, where the mean area
of a quasispherical cluster of fixed volume is shown to scale exactly
as $V^{2/3}$. Since $\alpha^*$ is roughly equal to the value of
$\langle{\cal A}\rangle_V/S-1$ for $V=V^*$, we expect the same
asymptotic behavior for both quantities.

%
%
\begin{figure}
\centering
\includegraphics[width=11cm]{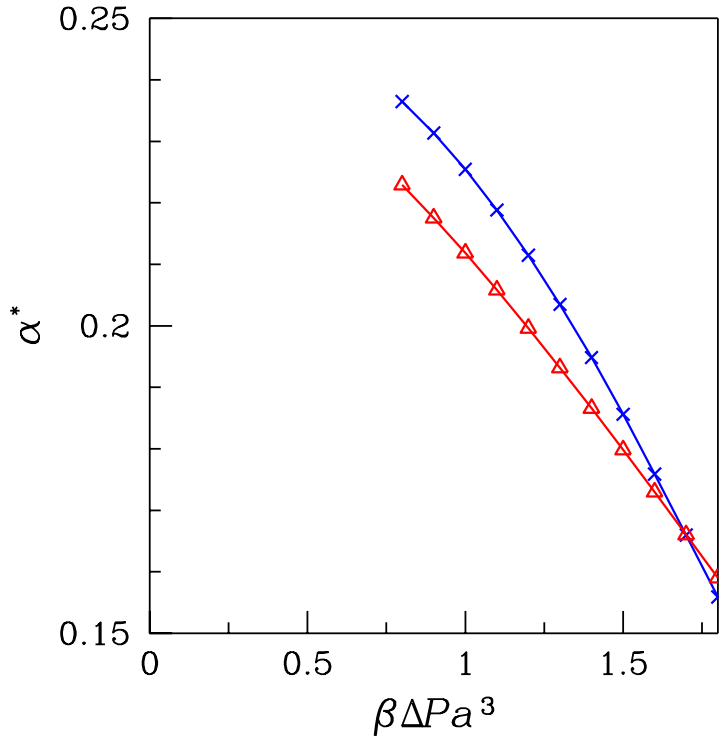}
\caption{(Color online). Quasispherical cluster: saddle-point value of
$\alpha$, plotted as a function of supersaturation, for
$\beta\sigma_m a^2=1$ and $\beta\lambda=0.2$
(blue crosses: $\delta_m=0.1\,a$; red triangles: $\delta_m=-0.1\,a$).
Approaching coexistence, where surface roughening ripples diverge, 
the ratio of the area of critical clusters to that of the equivalent sphere
remains finite ($\approx 1.25$).}
\label{fig3}
\end{figure}

\section{Extracting the interface tension from the nucleation rate}

Finally we consider whether employing one ($V$) or two CVs
($V$ and $A$) could affect the time-honored CNT extraction procedure
of the interface tension at coexistence, $\sigma_\infty$,
from the rate of nucleation $I$. Assuming the standard transition-state-theory
(Arrhenius-like) expression of $I$ for all supersaturations, i.e.,
$I=I_0\exp\{-\beta\Delta\Omega^*\}$, the most important source of
$I$ dependence on $\Delta P$ is the exponent, $-\beta\Delta\Omega^*$.
The latter quantity is plotted in Fig.\,4 for both one- and
two-CV cases, and for two different choices of parameters.
We point out that the near-coexistence slope of $-\beta\Delta\Omega^*$
is expected to be the same for both one- and two-dimensional surface free
energy, see our argument in appendix C.

%
%
\begin{figure}
\centering
\includegraphics[width=11cm]{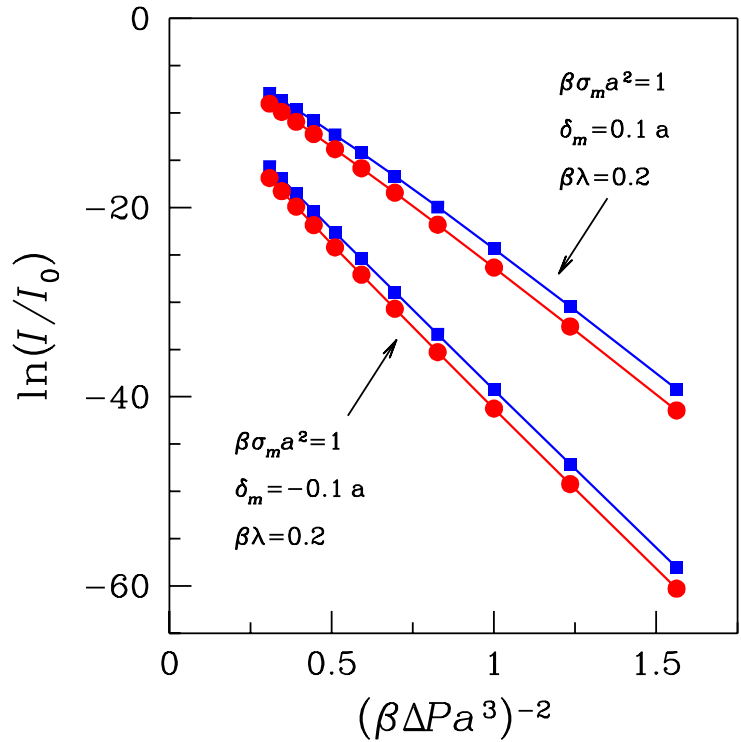}
\caption{(Color online).
Quasispherical cluster for $\beta\sigma_ma^2=1$ and $\beta\lambda=0.2$,
and for two opposite values of $\delta_m$. We plot $-\beta\Delta\Omega^*$
as a function of $(\Delta P)^{-2}$, which represents the leading $\Delta P$
dependence of $\ln I$ (blue squares: one-CV case; red dots: two-CV case).
The slope of $\ln(I/I_0)$
is nearly constant (i.e., CNT-like) only for very low supersaturations.
In this limit the slope of $\ln(I/I_0)$ appears to be the same for both
one- and two-CV cases (see text and appendix C, where a proof of
this equivalence is provided).
We observe that the direction of bending of $\ln I$ as a function of
$(\Delta P)^{-2}$ is a reliable marker of the sign of the Tolman length.}
\label{fig4}
\end{figure}

According to CNT, $\ln(I/I_0)$ should be a linear function of
$(\Delta P)^{-2}$, with a slope proportional to $\sigma_\infty^3$.
In the fluctuating-shape cluster model instead, $\ln(I/I_0)$ is a
concave function of $(\Delta P)^{-2}$ ($\delta_m>0$) or a convex
one ($\delta_m<0$), with the latter case apparently
applying for colloids (see e.g. Fig.\,7(b) of Ref.\,\cite{Franke}).
Hence, as was underlined in Ref.\,\cite{Prestipino1}, the correct
procedure of extracting the interface tension at
coexistence entails by necessity an extrapolation of the slope of
$\ln(I/I_0)$ at vanishing undercooling,
independently of whether we consider only $V$ or $(V, A)$ as CVs.

Quantitatively, the rate of nucleation is sensitive to the number of CVs
employed in the calculation: with two variables instead of one, $I$
is reduced by a few orders of magnitude for low supersaturations.

Since the limiting slope of $\ln I$ is the same for both one and two CVs,
a one-CV description of nucleation is sufficient
when the only objective is to get $\sigma_\infty$ out of a model of the
nucleation cluster. It is useful here to restate that the ``thermodynamical''
(i.e., dressed by thermal fluctuations)
surface tension $\sigma_\infty$, rather than the ``mechanical'' surface
tension $\tau$ (see appendix B), is what one obtains from a
measurement of the nucleation rate.

\section{Conclusions}

In nucleation, the minimum free-energy cost $\Delta\Omega$ for 
making a cluster of the stable phase (e.g. solid) out of the metastable 
parent phase (e.g. liquid) is the sum of two terms: a negative volume 
term, representing the benefit for switching a region from liquid to solid, 
and a positive surface term, $F_s$, which is the cost for creating
the interface. A crucial assumption of standard nucleation theories
is that the surface free energy $F_s$ only depends on $V$, the cluster
volume; at the critical size,
the reversible work of cluster formation reaches a maximum value,
which in turn determines the steady-state nucleation rate for low
enough undercooling.

Refining the standard description of the free energy of nucleation,
we have extended the theory of Refs.\,\cite{Prestipino1,Prestipino2}
using the area $A$ of the cluster surface as a second collective
variable besides volume $V$. The transition state is now a saddle point
in the two-dimensional free-energy surface, and the shape of the nucleus
is that of a corrugated sphere whose area relative to the equivalent sphere
depends upon 
the model parameters. We found that the inclusion of area systematically
corrects the barrier height upwards by a few $k_BT$, which in relative
terms may be important especially for low barriers.
Otherwise, the extrapolation procedure towards coexistence required
to extract the interface tension from the nucleation rate remains exactly
the same as for the volume-only case. In closing, we also
speculate that the effective rugosity, here signaled by the parameter $\alpha$,
might be expected to play a role in modifying the effective Stokes 
frictional force felt, e.g., by a solid nucleation cluster drifting in a fluid flow.

\section*{Acknowledgements}

This project was co-sponsored by 
the Italian Ministry of Education and Research through 
Contract PRIN/COFIN 2010LLKJBX\_004, and by ERC Advanced Grant
320796 MODPHYSFRICT. It also benefitted from the research environment 
and stimulus provided by SNF Sinergia Contract CRSII2 136287.

\appendix
\section{Calculation of $\Delta\Omega(V_2,A)$}
\setcounter{equation}{0}
\renewcommand{\theequation}{A.\arabic{equation}}

We here consider in more detail the calculation of $\Delta\Omega(V,A)$ 
for the case $l_{\rm max}=2$ (corresponding to $S/a^2=9$ and
$V/a^3=9/(2\sqrt{\pi})=2.53885\ldots$), which is perhaps the only case 
allowing for an analytic treatment. Assuming $c_2>0$, we
should compute the following integral:
\be
\int_{-\infty}^{+\infty}{\rm d}x\,\frac{e^{-i\alpha x}}{(c_2-2ix)^{5/2}}=c_2^{-3/2}\int_{-\infty}^{+\infty}{\rm d}x\,\frac{e^{-ic_2\alpha\,x}}{(1-2ix)^{5/2}}\equiv c_2^{-3/2}I(c_2\alpha)\,,
\label{a-1}
\ee
where
\be
I(\alpha)=\int_{-\infty}^{+\infty}{\rm d}x\,\frac{e^{-i\alpha x}}{(1-2ix)^{5/2}}\,.
\label{a-2}
\ee
The integrand is a complex function of real variable which does not show 
singularities on the integration path. Integrating twice by parts, we obtain:
\be
I=\frac{\alpha^2}{3}\int_{-\infty}^{+\infty}{\rm d}x\,\frac{e^{-i\alpha x}}{(1-2ix)^{1/2}}\,.
\label{a-3}
\ee
In order to determine (\ref{a-3}), we consider the complex integral
\be
\oint_\Gamma{\rm d}z\,\frac{e^{-i\alpha z}}{(1-2iz)^{1/2}}
\label{a-4}
\ee
over a keyhole circuit $\Gamma$ of the complex plane, see
Fig.\,5. The 
circuit is so chosen as to avoid the singularity of the integrand at
$z_0=-i/2$. Since there are no poles inside $\Gamma$, the integral 
(\ref{a-4}) simply vanishes. On the other hand, the same integral is the 
sum of various contributions, one of which approaches $I$ in the
$L\rightarrow+\infty$ limit.

%
%
\begin{figure}
\centering
\includegraphics[width=11cm]{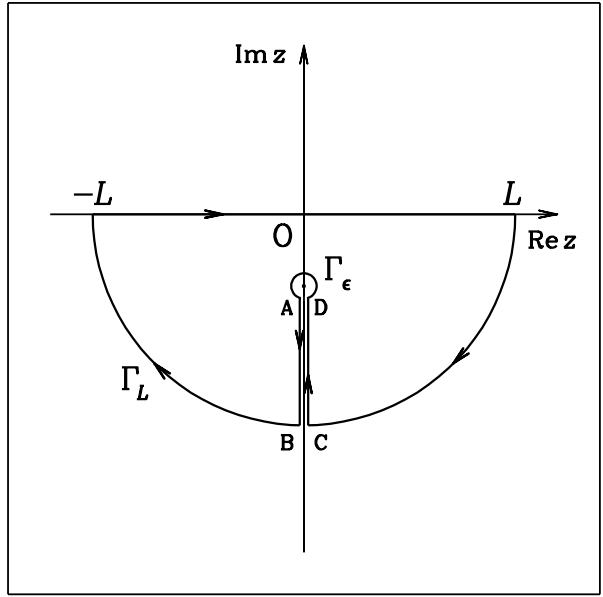}
\caption{(Color online). The integration path $\Gamma$ that was considered
in the evaluation of the integral (\ref{a-4}).}
\label{fig5}
\end{figure}

Let $\Gamma_L$ denote the semicircumference with center in the origin 
and radius $L$, lying in the half-plane ${\rm Im}z<0$, and
$\Gamma_\epsilon$ the circumference of radius $\epsilon$, centered in 
$z_0$. We shall prove later that the integrals over $\Gamma_L$ and
$\Gamma_\epsilon$ both vanish, respectively in the $L\rightarrow +\infty$ 
and $\epsilon\rightarrow 0^+$ limits. As far as the integrals over the 
segments $AB$ and $CD$ of Fig.\,5 are concerned,
they are given by
\be
\int_{z_0-\delta}^{-iL-\delta}{\rm d}z\,\frac{e^{-i\alpha z}}{(1-2iz)^{1/2}}\,\,\,\,\,\,{\rm and}\,\,\,\,\,\,\int_{-iL+\delta}^{z_0+\delta}{\rm d}z\,\frac{e^{-i\alpha z}}{(1-2iz)^{1/2}}\,,
\label{a-5}
\ee
$\delta$ being a small positive number.

Putting the branch cut of $z^{1/2}$ on the semiaxis of negative reals,
\be
z^{1/2}=\exp\left\{\frac{1}{2}\ln_{[-\pi,\pi)}z\right\}=\exp\left\{\frac{1}{2}\left(\ln|z|+i\arg_{[-\pi,\pi)}z\right)\right\}\,,
\label{a-6}
\ee
and taken $z=-i/2-it-\delta$, the first integral (\ref{a-5}) becomes
\be
-ie^{-\alpha/2}\int_0^{L-1/2}{\rm d}t\,\frac{e^{-\alpha t}e^{i\alpha\delta}}{\left(-2t+2i\delta\right)^{1/2}}\,,
\label{a-7}
\ee
where
\be
\left(-2t+2i\delta\right)^{1/2}=\exp\left\{\frac{1}{2}\left(\ln|-2t+2i\delta|+\pi i\right)\right\}\,\,\shortstack{\scriptsize $\delta\rightarrow 0^+$ \\ $\longrightarrow$}\,\,i(2t)^{1/2}\,.
\label{a-8}
\ee
Hence:
\be
\int_{z_0-\delta}^{-iL-\delta}{\rm d}z\,\frac{e^{-i\alpha z}}{(1-2iz)^{1/2}}\,\,\shortstack{\scriptsize $\delta\rightarrow 0^+$ \\ $\longrightarrow$}\,\,-e^{-\alpha/2}\int_0^{L-1/2}{\rm d}t\,\frac{e^{-\alpha t}}{(2t)^{1/2}}\,.
\label{a-9}
\ee
Similarly, since
\be
\left(-2t-2i\delta\right)^{1/2}=\exp\left\{\frac{1}{2}\left(\ln|-2t-2i\delta|-\pi i\right)\right\}\,\,\shortstack{\scriptsize $\delta\rightarrow 0^+$ \\ $\longrightarrow$}\,\,-i(2t)^{1/2}\,,
\label{a-10}
\ee
we find
\be
\int_{-iL+\delta}^{z_0+\delta}{\rm d}z\,\frac{e^{-i\alpha z}}{(1-2iz)^{1/2}}\,\,\shortstack{\scriptsize $\delta\rightarrow 0^+$ \\ $\longrightarrow$}\,\,e^{-\alpha/2}\int_{L-1/2}^0{\rm d}t\,\frac{e^{-\alpha t}}{(2t)^{1/2}}\,,
\label{a-11}
\ee
which is the same as (\ref{a-9}). After letting $L\rightarrow +\infty$, we 
finally obtain
\be
I=\frac{\alpha^2}{3}\sqrt{2}e^{-\alpha/2}\int_0^{+\infty}{\rm d}t\,\frac{e^{-\alpha t}}{\sqrt{t}}=\frac{1}{3}\alpha\sqrt{2\pi\alpha}\,e^{-\alpha/2}\,.
\label{a-12}
\ee
It remains to prove that the integrals over $\Gamma_L$ and
$\Gamma_\epsilon$ are irrelevant. As far as the former is concerned, it 
suffices to observe that its modulus is bounded from above by
\be
\frac{L}{(2L-1)^{1/2}}\int_\pi^{2\pi}{\rm d}\theta\,e^{L\alpha\sin\theta}<\frac{2L}{(2L-1)^{1/2}}\int_\pi^{3\pi/2}{\rm d}\theta\,e^{2L\alpha(1-\theta/\pi)}=\frac{\pi/\alpha}{(2L-1)^{1/2}}\left(1-e^{-L\alpha}\right)\,,
\label{a-13}
\ee
where we used the inequality
\be
\sin\theta<\frac{2}{\pi}(\pi-\theta)\,,
\label{a-14}
\ee
valid for $\pi<\theta<3\pi/2$. Moreover, we have
\be
\left|\int_{\Gamma_\epsilon}\frac{e^{-i\alpha x}}{(1-2ix)^{1/2}}\right|\le 2\pi\epsilon\,\frac{e^{-\alpha/2}}{(2\epsilon)^{1/2}}\,,
\label{a-15}
\ee
which vanishes as $\epsilon$ goes to zero.

We checked numerically that the result (\ref{a-12}) is correct by 
expressing $I$ in the equivalent form
\be
I=2\int_0^{+\infty}{\rm d}x\,\frac{\cos\left(\alpha x-\frac{5}{2}\arctan(2x)\right)}{(1+4x^2)^{5/4}}
\label{a-16}
\ee
and computing the integral numerically. Summing up, for $l_{\rm max}=2$ 
we obtain:
\be
\int_{-\infty}^{+\infty}{\rm d}x\,\frac{e^{-i\alpha x}}{(c_2-2ix)^{5/2}}=\frac{1}{3}\alpha\sqrt{2\pi\alpha}\,e^{-c_2\alpha/2}
\label{a-17}
\ee
and we get
\ba
e^{-\beta F_s(V_2,A)}&=&(36\pi)^{-2/3}\left(\frac{V_2}{a^3}\right)^{-4/3}e^{-8\pi\beta\lambda}e^{8\pi\beta\sigma_m\delta_m\left(3V_2/(4\pi)\right)^{1/3}}e^{-\beta\sigma_m A}\left(\frac{2\pi S_2}{s}\right)^8
\nonumber \\
&\times&\frac{1}{3\sqrt{2\pi}}\alpha^{3/2}\,e^{-c_2\alpha/2}\,,
\label{a-18}
\ea
with $S_2=9a^2,V_2=9a^3/(2\sqrt{\pi})$ and
$c_2=1+48\pi\beta\lambda-24\sqrt{\pi}\beta\sigma_m\delta_m a$.

It is now easy to check that Eq.\,(\ref{3-11}) is fulfilled for $l_{\rm max}=2$. 
From (\ref{a-18}) we get
\ba
e^{-\beta\Delta\Omega(V_2,A)} &=& (36\pi)^{-2/3}\left(\frac{V_2}{a^3}\right)^{-4/3}e^{-8\pi\beta\lambda}e^{8\pi\beta\sigma_m\delta_m\left(3V_2/(4\pi)\right)^{1/3}}e^{V_2\beta\Delta P-\beta\sigma_m S_2}
\nonumber \\
&\times& \left(\frac{2\pi S_2}{s}\right)^8\frac{1}{3\sqrt{2\pi}}\alpha^{3/2}\,e^{-(\alpha/2)\left(1+48\pi\beta\lambda-24\sqrt{\pi}\beta\sigma_m\delta_m a+2\beta\sigma_m S_2\right)}\,,
\label{a-19}
\ea
and we obtain
\ba
&& (36\pi)^{1/3}\left(\frac{V_2}{a^3}\right)^{2/3}\int_0^{+\infty}{\rm d}\alpha\,e^{-\beta\Delta\Omega(V_2,A)}=
\nonumber \\
&& (36\pi)^{-1/3}\left(\frac{V_2}{a^3}\right)^{-2/3}e^{-8\pi\beta\lambda}e^{8\pi\beta\sigma_m\delta_m\left(3V_2/(4\pi)\right)^{1/3}}e^{V_2\beta\Delta P-\beta\sigma_m S_2}
\nonumber \\
&\times& \left(\frac{2\pi S_2}{s}\right)^8\frac{1}{3\sqrt{2\pi}}\int_0^{+\infty}{\rm d}\alpha\,\alpha^{3/2}\,e^{-(\alpha/2)\left(1+48\pi\beta\lambda-24\sqrt{\pi}\beta\sigma_m\delta_m a+2\beta\sigma_m S_2\right)}\,.
\label{a-20}
\ea
Since
\be
\int_0^{+\infty}{\rm d}x\,x^{3/2}\,e^{-Kx}=\frac{3\sqrt{\pi}}{4}K^{-5/2}\,,
\label{a-21}
\ee
the final result is
\ba
&& \int_0^{+\infty}\frac{{\rm d}A}{a^2}\,e^{-\beta\Delta\Omega(V_2,A)}=(36\pi)^{-1/3}\left(\frac{V_2}{a^3}\right)^{-2/3}e^{-8\pi\beta\lambda}e^{8\pi\beta\sigma_m\delta_m\left(3V_2/(4\pi)\right)^{1/3}}
\nonumber \\
&\times& e^{V_2\beta\Delta P-\beta\sigma_m S_2}\left(\frac{2\pi S_2}{s}\right)^8\frac{1}{\left(1+48\pi\beta\lambda-24\sqrt{\pi}\beta\sigma_m\delta_m a+2\beta\sigma_m S_2\right)^{5/2}}\,,
\label{a-22}
\ea
which indeed is the value of $\exp\{-\beta\Delta\Omega(V_2)\}$ (cf.
Eq.\,(C16) of \cite{Prestipino2}; note that, due to an oversight, a term
$\exp\{\beta\rho_s|\Delta\mu|V\}$ was erroneously included in the 
expression of $Z_s$).

There is a more elegant way to obtain Eq.\,(\ref{a-18}). Upon rewriting
Eq.\,(\ref{3-12}) as
\ba
&& e^{-\beta F_s(V,A)}=(36\pi)^{-2/3}\left(\frac{V}{a^3}\right)^{-4/3}e^{-8\pi\beta\lambda}e^{8\pi\beta\sigma_m\delta_m\left(3V/(4\pi)\right)^{1/3}}e^{-\beta\sigma_m A}\left(\frac{2\pi S}{s}\right)^3
\nonumber \\
&\times& \int_{-\infty}^{+\infty}\prod_{l>1,m}\left(\frac{S}{s}{\rm d}x_{l,m}\right)
\exp\left(-\frac{1}{4\pi}\sum_{l>1,m}c_l(S)x_{l,m}^2\right)\delta\left(\frac{1}{8\pi}\sum_{l>1,m}(l^2+l-2)x_{l,m}^2-\alpha\right)\,,
\nonumber \\
\label{a-23}
\ea
one observes that, by a rescaling of the integration variables, the integral
in (\ref{a-23}) is converted to an integral over the
surface of a $M$-dimensional hypersphere, with $M=(l_{\rm max}+1)^2-4$.
We readily obtain:
\ba
&& e^{-\beta F_s(V,A)}=(36\pi)^{-2/3}\left(\frac{V}{a^3}\right)^{-4/3}e^{-8\pi\beta\lambda}e^{8\pi\beta\sigma_m\delta_m\left(3V/(4\pi)\right)^{1/3}}e^{-\beta\sigma_m A}\left(\frac{2\pi S}{s}\right)^3
\nonumber \\
&\times& \prod_{l=2}^{l_{\rm max}}\left(\frac{8\pi}{l^2+l-2}\right)^{l+1/2}\left(\frac{S}{s}\right)^{2l+1}\times\frac{1}{2\sqrt{\alpha}}\int_{S^M(\sqrt{\alpha})}{\rm d}S
\exp\left(-\sum_{l>1,m}\frac{2c_l}{l^2+l-2}x_{l,m}^2\right)\,,
\nonumber \\
\label{a-24}
\ea
where $S^M(\sqrt{\alpha})$ denotes the surface of the $M$-dimensional
hypersphere of radius $\sqrt{\alpha}$. The integral in (\ref{a-24}) is trivial
for $l_{\rm max}=2$, where we are again led to the result (\ref{a-18}).
For $l_{\rm max}>2$, the surface integral may still be evaluated numerically
by resorting to Monte Carlo sampling~\cite{Krauth}, but the computation is
feasible only when $l_{\rm max}$ is not too large.

\section{Mean area and width of a quasispherical cluster of fixed volume}
\setcounter{equation}{0}
\renewcommand{\theequation}{B.\arabic{equation}}

In this appendix, we establish a number of formulae for a quasispherical
cluster of fixed volume, which extend to spherical geometry known properties
of a rough planar interface.

For a quasispherical interface of volume $V$, statistical averages are
computed with a weight proportional to
$\exp\{-\beta{\cal H}_s\}\delta({\cal V}[\Sigma]-V)$ with ${\cal H}_s$
given by Eq.\,(\ref{2-3}). In particular, using Eq.\,(C2) of Ref.\,\cite{Prestipino2},
the average interface area reads
\be
\langle{\cal A}[\Sigma]\rangle_V=S\left(1+\frac{1}{4}\sum_{l=2}^{l_{\rm max}}\frac{(2l+1)(l-1)(l+2)}{b_l}\right)
\label{b-1}
\ee
with
\be
b_l=1+\frac{\beta\sigma_m}{2}S(l^2+l-2)-4\pi\beta\sigma_m\delta_m\left(\frac{S}{4\pi}\right)^{1/2}(l^2+l-2)+2\pi\beta\lambda\,l(l+1)(l-1)(l+2)\,.
\label{b-2}
\ee
The large-$V$ behavior of (\ref{b-1}) can be extracted by the
Euler-Mac Laurin formula, leading eventually to
\be
\frac{\langle{\cal A}[\Sigma]\rangle_V}{S}=1+\frac{k_BT}{8\pi\lambda}\ln\left(1+\frac{4\pi\lambda}{\sigma_m a^2}\right)+\frac{2\sqrt{\pi}\delta_m}{\beta\sigma_m a^2+4\pi\beta\lambda}\frac{1}{\sqrt{S}}+{\cal O}(S^{-1})
\label{b-3}
\ee
(for example, the asymptotic value of $\langle{\cal A}[\Sigma]\rangle_V/S$
for $\beta\sigma_m a^2=1$ and $\beta\lambda=0.2$ is 1.249982...).
A more elegant way to derive (\ref{b-3}) is to observe that, by
Eq.\,(\ref{3-5}),
\be
\langle{\cal A}[\Sigma]\rangle_V=\left.\frac{\partial F_s(V)}{\partial\sigma_m}\right|_{\sigma_m\delta_m}\,.
\label{b-4}
\ee
Using Eq.\,(C21) of Ref.\,\cite{Prestipino2}, we readily arrive at (\ref{b-3}).
We successfully checked Eq.\,({\ref{b-3}}) in a few cases
also by directly computing the sum in (\ref{b-1}). In particular,
$\langle{\cal A}[\Sigma]\rangle_V/S$ indeed approaches
its limiting value from above when $\delta_m>0$.

The result (\ref{b-3}) is akin to
\be
\frac{\left<{\cal A}\right>}{L^2}\sim 1+\frac{k_BT}{8\pi\lambda}\ln\left(1+\frac{\lambda\pi^2}{\gamma a^2}\right)\,,
\label{b-5}
\ee
which applies for a solid-on-solid interface with projected area $L^2$
and Hamiltonian
\be
{\cal H}[h]=\gamma L^2+\frac{1}{2}\int_D{\rm d}x\,{\rm d}y\left[\gamma(\nabla_\perp h)^2+\lambda(\nabla_\perp^2h)^2\right]\,,
\label{b-6}
\ee
where $\nabla_\perp=\partial_x\hat{\bf x}+\partial_y\hat{\bf y}$ and the integral
is extended over a square (the domain $D$) of area $L^2$. In Eq.\,(\ref{b-5})
the characteristic length $a$ arises from the lattice regularization of (\ref{b-6}),
which is a necessity if we are to avoid the divergence of the partition function. 
Observe that the Gaussian interface described by Eq.\,(\ref{b-6}) is always
rough, since
\be
\left<(h_{\bf x}-h_{{\bf x}'})^2\right>\sim\frac{k_BT}{\pi\gamma}\ln\frac{|{\bf x}-{\bf x}'|}{a}\,.
\label{b-7}
\ee
This latter result is easily translated to the sphere, by observing that
the average square width of a quasispherical cluster reads
\be
\frac{1}{4\pi}\int{\rm d}^2\Omega\,\langle(R(\theta,\phi)-R_0)^2\rangle_V=\frac{S}{2}\sum_{l=1}^{l_{\rm max}}\frac{2l^2+l+1}{(2l+1)b_l}\sim\frac{k_BT}{3\sigma_m}\ln\left(\frac{V}{a^3}\right)\,.
\label{b-8}
\ee
Besides certifying that a quasispherical interface is technically rough,
Eq.\,(\ref{b-8}) also indicates that the average size of the deviation
$R/R_0-1=\epsilon$ from sphericity scales as $\sqrt{S^{-1}\ln(S/a^2)}$ for
large clusters; on the other hand, for the smallest clusters the angular
average of $\langle\epsilon^2\rangle_{V}^{1/2}$ can be as large as
$0.6$ for typical values of the model parameters. Hence we confirm
that the quasispherical model is a near-coexistence approximation
only rigorously valid for small to moderate undercooling.

We note in passing that the average cluster area can be put in relation
with the {\em mechanical} surface tension $\tau$~\cite{Imparato}, which
measures the elastic response of an interface to a change in its projected
area. In the planar case (\ref{b-6}), the stretching or shrinking of the projected
area is obtained by changing $a$, the lattice spacing, at fixed number $N$
of lattice sites. The result is
\be
\tau\equiv\frac{1}{N}\frac{\partial F_s}{\partial(a^2)}=\gamma\frac{\left<{\cal A}\right>}{L^2}-\frac{k_BT}{2a^2}\,.
\label{b-9}
\ee
Similarly, in the quasispherical case $\tau$ can be obtained by keeping the
total number of $(l,m)$ modes fixed while differentiating $F_s(V)$ with
respect to $S$. Calling $N=(l_{\rm max}+1)^2=S/a^2$, we first rewrite
$F_s$ as (see Eq.\,(C18) in Ref.~\cite{Prestipino2})
\ba
F_s&=&-2k_BT\ln N+\sigma_m Na^2+8\pi\lambda-8\pi\sigma_m\delta_m\left(\frac{Na^2}{4\pi}\right)^{1/2}+3k_BT\ln 2
\nonumber \\
&+&\frac{k_BT}{2}\sum_{l=2}^{l_{\rm max}}(2l+1)\left(-2\ln\frac{N}{2}+\ln b_l\right)\,.
\label{b-10}
\ea
After simple algebra, we get
\be
\tau=\sigma_m\left(1-\delta_m\left(\frac{4\pi}{S}\right)^{1/2}\right)\frac{\langle{\cal A}[\Sigma]\rangle_V}{S}\,,
\label{b-11}
\ee
which nicely recalls Eq.\,(\ref{b-9}).

\section{Large-size limit of the two-dimensional surface free energy}
\setcounter{equation}{0}
\renewcommand{\theequation}{C.\arabic{equation}}

In Ref.\,\cite{Prestipino2}, the $V$-dependent surface free energy of
a quasispherical cluster was written as $F_s=\sigma(S)S$, where in
the large-size limit $\sigma(S)=\sigma(\infty)+{\cal O}(S^{-1/2})$
(see Eq.\,(C21) of Ref.\,\cite{Prestipino2}).
Similarly, we are here interested in establishing the behavior of
$F_s(V,A)$ in the limit where $V\rightarrow\infty$ for fixed
$\alpha=A/S-1$. A likely possibility, suggested by the profiles of
$F_s(V,\alpha)$ for increasing $V$ values (see Fig.\,6),
is that
\be
\beta F_s(V,\alpha)=f(\alpha)S+g(\alpha)o(S)\,,
\label{c-1}
\ee
denoting $o(S)$ a quantity growing slower than $S$ for
$S\rightarrow\infty$ and $f,g$ two not further specified functions of
$\alpha$. Figure 6 indicates that $f(\alpha)$ has a
minimum value, $f_{\rm min}=f(\alpha_{\rm min})$, falling not far
away from the asymptotic value of
$\langle{\cal A}[\Sigma]\rangle_V/S-1$. By the same $F_s$
data reported in Fig.\,6 we infer that the first
subdominant term in (\ref{c-1}) is actually a $\sqrt{S}$ term.

%
%
\begin{figure}
\centering
\includegraphics[width=11cm]{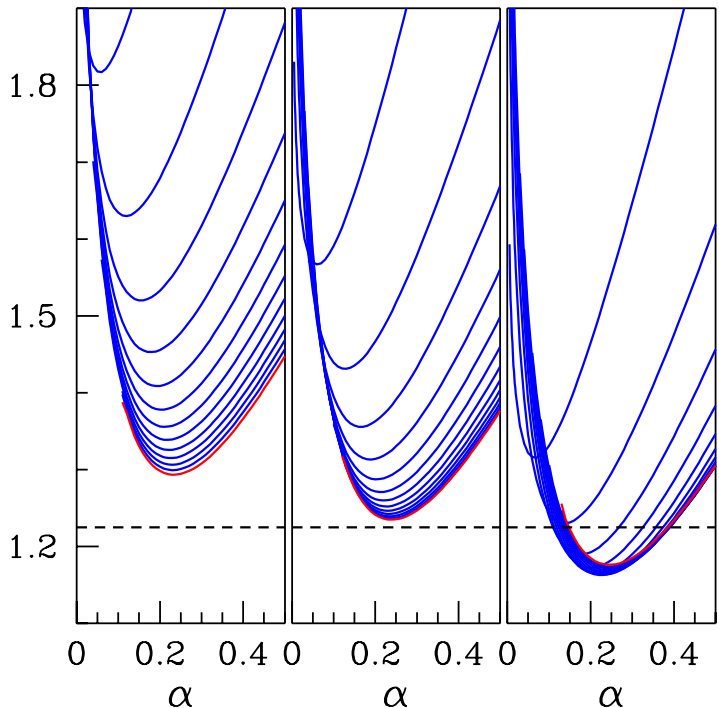}
\caption{(Color online). Ratio of $\beta F_s(V,\alpha)$ to $S$ for
increasing $V$ values ($V/a^3=(l_{\rm max}+1)^3/\sqrt{36\pi}$
with $l_{\rm max}=2,\ldots,14$). Low $V$ values are on top, and
the red curve refers to $l_{\rm max}=14$. Three sets of parameters
were investigated: $\beta\sigma_m a^2=1,\beta\lambda=0.2$, and
$\delta_m=-0.1,0,0.1$ (from left to right). Upon increasing $V$,
$\beta F_s(V,\alpha)/S$ approaches a limiting profile whose
minimum coincides with $\beta\sigma(\infty)$ (the dashed
line). For $\delta_m>0$, the approach to this limit is non-monotonic.}
\label{fig6}
\end{figure}

Assuming that (\ref{c-1}) holds, we now prove that
$f_{\rm min}=\beta\sigma(\infty)$.
Starting from Eq.\,(3.11), which we reshuffle as
\be
-\ln\int_S^{+\infty}\frac{{\rm d}A}{a^2}\,e^{-\beta F_s(V,A)}=\beta F_s(V)\,,
\label{c-2}
\ee
we divide each side of (\ref{c-2}) by $S$ and then bring the volume to infinity:
\be
-\lim_{V\rightarrow\infty}\frac{1}{S}\ln\int_0^{+\infty}{\rm d}\alpha\,\frac{S}{a^2}e^{-\beta F_s(V,\alpha)}=\beta\sigma(\infty)\,.
\label{c-3}
\ee
Upon carrying the limit inside the integral (which is allowed in so far as
$\alpha$ is independent of $V$), the left-hand side of (\ref{c-3}) becomes
\be
-\lim_{V\rightarrow\infty}\frac{1}{S}\ln\int_0^{+\infty}{\rm d}\alpha\,e^{-f(\alpha)S-g(\alpha)o(S)}\,,
\label{c-4}
\ee
in turn equal to $f_{\rm min}$ by the Laplace (saddle-point) method.
Alternatively, we may also expand for large $S$ both $f$ and $g$ to
second-order in the deviation of $\alpha$ from $\alpha_{\rm min}$.
By matching the two sides of Eq.\,(\ref{c-2}), we again find
$f_{\rm min}=\beta\sigma(\infty)$ and moreover (by Eq.\,(C21)
of Ref.\,\cite{Prestipino2})
\be
g(\alpha_{\rm min})=-2\delta_m\sqrt{\pi}\left[2\beta\sigma_m+\frac{\sigma_m}{4\pi\lambda}\ln\left(1+\frac{4\pi\lambda}{\sigma_m a^2}\right)\right]\,.
\label{c-5}
\ee

The above result can be taken as the
proof that the slope of $-\beta\Delta\Omega^*$ at vanishing undercooling
is the same for both one and two-CV descriptions of nucleation.
In fact, let it be assumed that $\Delta P$ is so low that we are authorized
to take $\beta F_s(V,\alpha)=f(\alpha)S$. Then, the extremal point
(saddle point) of $\Delta\Omega(V,\alpha)=-V\Delta P+F_s(V,\alpha)$
is the unique solution to
\be
f'(\alpha)=0\,\,\,\,\,\,{\rm and}\,\,\,\,\,\,-\beta\Delta P+\frac{2}{3}(36\pi)^{1/3}f(\alpha)V^{-2/3}=0\,,
\label{c-6}
\ee
giving eventually
\be
R^*\equiv\left(\frac{3V^*}{4\pi}\right)^{1/3}=\frac{2\sigma(\infty)}{\Delta P}\,\,\,\,\,\,{\rm and}\,\,\,\,\,\,\Delta\Omega^*=\frac{16\pi}{3}\frac{\sigma(\infty)^3}{(\Delta P)^2}\,.
\label{c-7}
\ee
These values of critical radius and barrier height are the same occurring
in CNT when the interface tension is chosen to be $\sigma(\infty)$.


\begin{thebibliography}{99}
\bibitem{Kashchiev} See, for example, D. Kashchiev, {\em Nucleation: 
Basic Theory with Applications} (Butterworth-Heinemann, Oxford, 2000).

\bibitem{Prestipino1} S. Prestipino, A. Laio, and E. Tosatti, {\em Phys. Rev. 
Lett.} {\bf 108}, 225701 (2012).

\bibitem{Prestipino2} S. Prestipino, A. Laio, and E. Tosatti, {\em J. Chem. 
Phys.} {\bf 138}, 064508 (2013).

\bibitem{Fisher} M. P. A. Fisher and M. Wortis, {\em Phys. Rev. B} {\bf 29}, 
6252 (1984).

\bibitem{Pan} A. C. Pan and D. Chandler,
{\em J. Phys. Chem. B} {\bf 108}, 19681 (2004).

\bibitem{Moroni} D. Moroni, P. R. ten Wolde, and P. G. Bolhuis, {\em 
Phys. Rev. Lett.} {\bf 94}, 235703 (2005).

\bibitem{Trudu} F. Trudu, D. Donadio, and M. Parrinello, {\em Phys. Rev. 
Lett.} {\bf 97}, 105701 (2006).

\bibitem{Peters} B. Peters and B. L. Trout, {\em J. Chem. Phys.} {\bf 125}, 
054108 (2006).

\bibitem{Zykova-Timan}  T. Zykova-Timan, C. Valeriani, E. Sanz, D. Frenkel, 
and E. Tosatti, {\em Phys. Rev. Lett.} {\bf 100}, 036103 (2008).

\bibitem{Lechner} W. Lechner, C. Dellago, and P. G. Bolhuis, {\em Phys. 
Rev. Lett.} {\bf 106}, 085701 (2011).

\bibitem{Russo} J. Russo and H. Tanaka, {\em Scientific Reports} {\bf 2}, 
505 (2012).

\bibitem{tenWolde} P. R. ten Wolde and D. Frenkel, {\em J. Chem. Phys.} 
{\bf 109}, 9901 (1998).

\bibitem{Bowles} R. K. Bowles, R. McGraw, P. Schaaf, B. Senger, J.-C. 
Voegel, and H. Reiss, {\em J. Chem. Phys.} {\bf 113}, 4524 (2000).

\bibitem{Maibaum} L. Maibaum, {\em Phys. Rev. Lett.} {\bf 101}, 019601 
(2008).

\bibitem{tenWolde2} P. R. ten Wolde, M. J. Ruiz-Montero, and D. Frenkel,
{\em Faraday Discuss.} {\bf 104}, 93 (1996).

\bibitem{Burkhardt} T. Burkhardt, {\em J. Phys. A} {\bf 14}, L63 (1981).

\bibitem{Franke} M. Franke, A. Lederer, and H. J. Sch\"{o}pe,
{\em Soft Matter} {\bf 7}, 11267 (2011).

\bibitem{Krauth} W. Krauth, {\em Statistical Mechanics: Algorithms and
Computations} (Oxford University Press, Oxford, 2006).

\bibitem{Imparato} A. Imparato, {\em J. Chem. Phys.} {\bf 124}, 154714
(2006).
\end{thebibliography}
\end{document}